\documentclass[aps,pra,twocolumn,floatfix,superscriptaddress]{revtex4}

\usepackage{amssymb,amsmath,amstext}                
\usepackage{graphicx}                                               
\usepackage{epstopdf}                                               
\usepackage{color}                                                     
\usepackage{bm}                                                        
\usepackage{appendix}                                              
\usepackage{bbold}
\usepackage{bbm}
\usepackage{ulem}
\usepackage{latexsym}
\usepackage[colorlinks=true,citecolor=blue,linkcolor=magenta]{hyperref}
\usepackage{cancel}

\def\be{\begin{equation}}
\def\ee{\end{equation}}
\def\bea{\begin{eqnarray}}
\def\eea{\end{eqnarray}}
\def\ra{\rangle}
\def\la{\langle}
\def\bi{\begin{itemize}}
\def\ei{\end{itemize}}

\definecolor{dgreen} {RGB}{78,138,21}

\begin{document} 

\title{Dynamical quantum phase transitions in discrete time crystals}

\author{Arkadiusz Kosior} 
\affiliation{
Instytut Fizyki imienia Mariana Smoluchowskiego, 
Uniwersytet Jagiello\'nski, ulica Profesora Stanis\l{}awa \L{}ojasiewicza 11, PL-30-348 Krak\'ow, Poland}
\author{Krzysztof Sacha} 
\affiliation{
Instytut Fizyki imienia Mariana Smoluchowskiego, 
Uniwersytet Jagiello\'nski, ulica Profesora Stanis\l{}awa \L{}ojasiewicza 11, PL-30-348 Krak\'ow, Poland}
\affiliation{Mark Kac Complex Systems Research Center, Uniwersytet Jagiello\'nski, ulica Profesora Stanis\l{}awa \L{}ojasiewicza 11, PL-30-348 Krak\'ow, Poland
}

\date{\today}

\begin{abstract}
Discrete time crystals are related to non-equilibrium dynamics of periodically driven quantum many-body systems where the discrete time translation symmetry of the Hamiltonian is spontaneously broken into another discrete symmetry. Recently, the concept of phase transitions has been extended to non-equilibrium dynamics of time-independent systems induced by a quantum quench, i.e. a sudden change of some parameter of the Hamiltonian. There, the return probability of a system to the ground state reveals singularities in time which are dubbed dynamical quantum phase transitions. We show that the quantum quench in a discrete time crystal leads to dynamical quantum phase transitions where the return probability of a periodically driven system to a Floquet eigenstate before the quench reveals singularities in time. It indicates that dynamical quantum phase transitions are not restricted to time-independent systems and can be also observed in systems that are periodically driven.  We discuss how the phenomenon can be observed in ultra-cold atomic gases.
\end{abstract}
\date{\today}

\maketitle

\section{Introduction}

Phase transitions in equilibrium statistical physics are related to abrupt changes of macroscopic properties of many-body systems \cite{Sachdev2011,SachdevKeimer2011}. Macroscopic quantities that characterize systems in the thermodynamic limit reveal non-analytical behavior as a function of a control parameter. The equilibrium phase transitions are much better understood than non-equilibrium dynamics of quantum many-body systems \cite{Eckstein2009,Garrahan2010,Diehl2010,Schiro2010,Sciolla2010,Sciolla2011,Ates2012,Sciolla2013,Vosk2014,Zunkovic2015,Maraga2016}. Recently, it has been shown that real-time evolution of time-independent many-body systems, after a quantum quench, can reveal non-analytical behavior at a critical time \cite{Heyl2013,Heyl2015,Budich2016,Sharma2016,Bhattacharya2017a,Bhattacharya2017b,Karrasch2017,Heyl2017,Halimeh2017,Zauner-Stauber2017,Homrighausen2017,Lang2017,Zunkovic2016}. That is, starting with a system in the ground state, a sudden change of some parameter of the Hamiltonian results in non-analytical evolution of the return probability to the initial ground state. This phenomenon has been termed dynamical quantum phase transition and it has been already demonstrated in experiments \cite{Flaschner2016,Jurcevic2017}, for review see \cite{Heyl2017rev}. 

In 2012 Frank Wilczek suggested that periodic structures in time could be formed spontaneously by a quantum many-body system \cite{Wilczek2012}. The original idea of such a time crystal could not be realized because it assumed a system in the ground state \cite{Bruno2013b,Watanabe2015,Syrwid2017,Iemini2017,Huang2017a,Prokofev2017}. However, soon it turned out that periodically driven quantum many-body systems were able to self-re-organize their motion and spontaneously start evolving with a period which was different than a period of an external driving \cite{Sacha2015,Khemani16,ElseFTC,Yao2017,Lazarides2017,Russomanno2017,Zeng2017,Nakatsugawa2017,Ho2017,Huang2017,Gong2017,Wang2017}. These quantum phenomena are dubbed discrete time crystals because discrete time translation symmetry is broken into another discrete symmetry. Discrete time crystals have been recently observed experimentally \cite{Zhang2017,Choi2017,Nayak2017}. It should be stressed that in the classical regime breaking of discrete time translation symmetry in an atomic system has been also demonstrated in a laboratory \cite{Kim2006,Heo2010}. Wilczek idea initiated a new research area where non-trivial crystalline structures are investigated in the time domain \cite{Guo2013,Sacha15a,sacha16,Guo2016,Guo2016a,Giergiel2017,Mierzejewski2017,delande17,Flicker2017,Liang2017,Giergiel2017a}, for review see \cite{Sacha2017rev}.

Discrete time crystal formation takes place if interactions between particles are sufficiently strong. If they are not, exact many-body Floquet eigenstates evolve with a period of an external driving and are not vulnerable to infinitesimally weak perturbations. However, when the strength of particle interactions is greater than a critical value, Floquet eigenstates possess Schr\"odinger cat like structures and any perturbation, or even measurement of a position of a single particle, has a dramatic effect on system dynamics leading to a change of the period of motion \cite{Sacha2015,Sacha2017rev}. Here we will show that starting with a Floquet eigenstate in the regime of time crystal formation, an abrupt change of the particle interaction strength to the weak interaction regime induces dynamical quantum phase transitions. That is, the return probability to the initial Floquet eigenstate reveals singularities in time. In the second part of the article we  consider singularities in the dynamics of states with spontaneously broken time translation symmetry, and identify experimentally measurable  observables.

\section{Results}

We will focus on the discrete time crystal described in Ref.~\cite{Sacha2015}, i.e. on ultra-cold bosonic atoms bouncing on a harmonically oscillating (with frequency $\omega$) mirror in the presence of the gravitational force \cite{Steane95,Lau1999,Bongs1999,Buchleitner2002}. Let us begin with a single particle problem. In the frame moving with the mirror and in the gravitational units the Hamiltonian for a single-particle system reads $H_0=\frac{p^2}{2}+x+\Lambda x\cos(\omega t)$ where $x\ge 0$, i.e. the mirror is located at $x=0$ in the moving frame, and $\frac{\Lambda}{\omega^2}$ is the amplitude of the mirror oscillations in the laboratory frame \cite{Buchleitner2002}. Classical description of a single particle reveals resonant periodic orbits with periods equal to integer multiples of the driving period $\frac{2\pi}{\omega}$. We will concentrate on the 2:1 resonance where the classical resonant orbit possesses the period $2\frac{2\pi}{\omega}$. In the quantum description there exist two Floquet eigenstates which are represented by two orthogonal superpositions, $u_{1,2}(x,t)\propto \phi_1\pm\phi_2$, of two localized wavepackets, $\phi_{1,2}(x,t)$, that move along the 2:1 resonant orbit like a classical particle. Each of these two wavepackets evolves with the period $2\frac{2\pi}{\omega}$ but because after every period $\frac{2\pi}{\omega}$ they exchange their roles, the Floquet eigenstates $u_{1,2}$ are periodic with the period of the external driving. The two Floquet states are eigenstates of the single particle Floquet Hamiltonian, $(H_0-i\partial_t)u_{1,2}=\varepsilon_{1,2}u_{1,2}$, corresponding to quasi-energies $\varepsilon_2=\varepsilon_1+J$ (modulo $\frac{\omega}{2}$) where $J$ is an amplitude related to tunneling of a particle from one of the wavepacket to the other one. 

In order to find many-body Floquet eigenstates for ultra-cold atoms bouncing on the oscillating mirror we assume that interaction energy per particle is much smaller than the energy gap for excitation of the localized wavepackets. In the following we use the parameters as in Ref.~\cite{Sacha2015}, i.e. $\Lambda=0.06$ and $\omega=1.1$, then the energy gap is about $10^3J$ while the interaction energy we consider is of the order of $10J$. Therefore, we may restrict to the consideration of behavior of the $N$-body system in the Hilbert subspace spanned by Fock states $|N-n,n\ra$ where $N-n$ and $n$ are occupations of the localized wavepackets $\phi_1$ and $\phi_2$, respectively \cite{Sacha2015,Sacha15a}. This approximation resembles the two-mode approximation known in the description of a many-body system in a double-well potential \cite{Raghavan1999}. In the time-dependent two-mode basis $\{\phi_1(x,t),\phi_2(x,t)\}$, our $N$-body Floquet Hamiltonian reduces to 
\bea
\hat H&=&-\frac{J}{2}(\hat a_1^\dagger\hat a_2+\hat a_2^\dagger\hat a_1)+\frac{U}{2}(\hat a_1^\dagger\hat a_1^\dagger\hat a_1\hat a_1+\hat a_2^\dagger\hat a_2^\dagger\hat a_2\hat a_2) \cr
&&+2U_{12}\hat a_1^\dagger \hat a_1\hat a_2^\dagger \hat a_2,
\label{h}
\eea
where the standard bosonic operators $\hat a_{1,2}$ annihilate particles in the modes $\phi_{1,2}$, $U=g_0\int_0^{4\pi/\omega}dt\int_0^\infty dx |\phi_{1,2}|^4$ and $U_{12}=g_0\int_0^{4\pi/\omega}dt\int_0^\infty dx |\phi_{1}|^2|\phi_2|^2$ with $g_0$ determined by $s$-wave scattering length of atoms \cite{Sacha2015}. For very weak interactions, i.e. for $\gamma =N(U-2U_{12})/J\approx 0$, the ground state of the Hamiltonian (\ref{h}) corresponds to a Bose-Einstein condensate where all bosons occupy the single particle Floquet state $u_1(x,t)$. However, if particle interactions are attractive ($g_0<0$) and sufficiently strong, $\gamma\ll-1$, it is energetically favorable to group all atoms in one of the localized wavepackets. Then, for large $N$, low-lying eigenstates of the Hamiltonian (\ref{h}) are dominated by pairs of nearly degenerate Schr\"odinger cat like states \cite{Zin2008,Oles2010}. For example, the lowest two eigenstates of (\ref{h}) are $|\psi_\pm(t)\ra\approx(|N,0\ra\pm|0,N\ra)/\sqrt{2}$. Such many-body Floquet eigenstates evolve with the period of the external driving $\frac{2\pi}{\omega}$. However, even if the many-body system is prepared initially in one of such Schr\"odinger cat like states, e.g. in the Floquet eigenstate $|\psi_+\ra$, measurement of a position of a single particle leads to a collapse of the eigenstate to $|N,0\ra$ or $|0,N\ra$ state what breaks the original time translation symmetry because the subsequent time evolution takes place with the period  $2\frac{2\pi}{\omega}$ \cite{Sacha2015}. The lifetime of the symmetry broken state goes to infinity when $N\rightarrow\infty$ but $\gamma=$constant.

In the first part  of this article we do not consider spontaneous breaking of time translation symmetry but we assume that the perfectly isolated system is prepared in the Floquet eigenstate $|\psi_+(t)\ra=K(t,0)|\psi_+(0)\ra$, where $K(t,0)$ is the time evolution operator corresponding to $\gamma<-1$ and $|\psi_+(0)\ra$ is the Floquet eigenstate at $t=0$. Subsequently, we assume that at $t=t_0>0$ the $s$-wave scattering length $g_0$ is suddenly changed to, e.g., zero. After the quench, the state evolves according to the new time evolution operator: $|\tilde{\psi}_+(t)\ra=\tilde{K}(t,t_0)|\psi_+(t_0)\ra$. Time evolution of $|\tilde{\psi}_+(t)\ra$ can be easily obtained by numerical integration of the many-body Schr\"odinger equation with the Hamiltonian (\ref{h}) (this Hamiltonian can be rewritten in the form of the spin system Hamiltonian \cite{Milburn1997} or the infinite-range Ising model \cite{Zunkovic2015}). However, it is much more instructive to apply the so-called continuum approximation \cite{Zin2008} which reduces the many-body Hamiltonian (\ref{h}) to the Hamiltonian of a fictitious particle, 
\be
\hat H=-\frac{J}{N}\sqrt{1-z^2}\frac{\partial^2}{\partial{z^2}}+V(z), 
\label{ch}
\ee
in the presence of the effective potential,
\be
V(z)=\frac{JN}{2}\left(-\sqrt{1-z^2}+\frac{\gamma}{2}z^2\right).
\label{effp}
\ee
Wavefunction $\psi(z)$ of the fictitious particle is a many-body state written in the Fock space basis $|N-n,n\ra$ but with the assumption that number of particles $N$ is so large that the relative population difference $z_n=\frac{(N-n)-n}{N}$ can be treated as a continuous variable $z$ with the restriction $|z|\le 1$. 

There are two crucial ranges of the parameter $\gamma$. For $\gamma<-1$ the effective potential (\ref{effp}) possesses a double well structure, while for $\gamma>-1$ it has a single well shape \cite{Zin2008}. For $\gamma<-1$, the ground state of the fictitious particle can be approximated by the superposition $\psi_+(z) = [\psi_L(z)+\psi_R(z)]/\sqrt{2}$, where $\psi_{L,R}$ are Gaussian states. (See the Appendix A for a derivation of the continuum approximation, and the analytical solutions for  $|\tilde{\psi}_+(t)\ra$.)
The state $\psi_+(z)$ is the previously described $N$-body Floquet eigenstate $|\psi_+(t)\ra = \sum_n \psi_+(z_n)|N-n,n\ra$ written in the time-dependent Fock basis and within the continuum approximation. 

\begin{figure}[tb] 	            
\includegraphics[width=0.95\columnwidth]{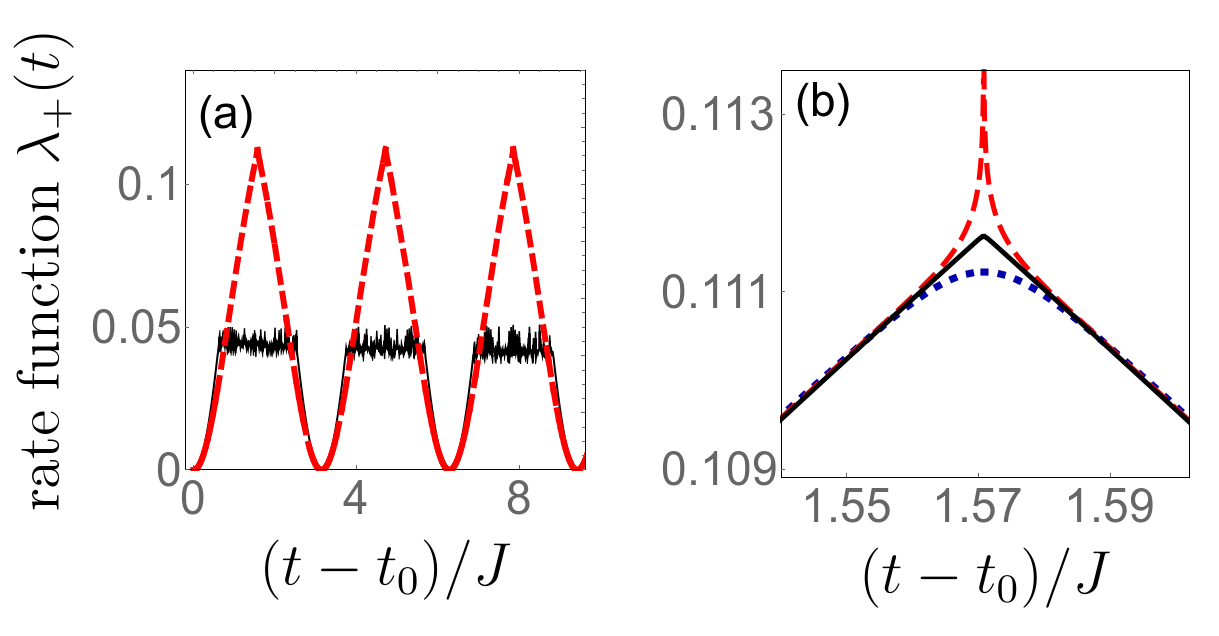}       
\caption{Quench from $\gamma=-1.15$ to $\gamma=0$ in the system prepared initially in the Floquet state $|\psi_+(t)\ra$.
Panel (a): the rate function $\lambda_+^{(N)}(t)$ obtained numerically (solid black line) and analytically within the continuum approximation (red dashed line) for $N=1500$. Both curves follow each other except time intervals when the Loschmidt echo is so small that the numerical precision breaks down. Panel (b): rate function $\lambda_+^{(N)}(t)$ for $N=$2000 (blue dotted), 4869 (red dashed) and 50000 (black solid) 
in the vicinity of the critical time $t_c=t_0+\frac{\pi}{2J}$. The diverging curve is related to an accidental value of $N$ when the Loschmidt echo vanishes, cf. (\ref{lambda_tc}). 
}
\label{Loschmidt}   
\end{figure}

Let us assume that the system is prepared in the ground state $\psi(z)=\psi_+(z)$  of the Hamiltonian (\ref{ch}) for $\gamma<-1$ and at $t=t_0$ we suddenly set the scattering length $g_0$ to zero, i.e. we switch to $\gamma=0$. The state $\tilde\psi(z,t>t_0)$ evolves according to the new Hamiltonian and we are interesting in the so-called Loschmidt echo \cite{Heyl2013}, i.e. the return probability to the initial state, $|{\cal G}(t>t_0)|^2$, where ${\cal G}(t)=\int_{-1}^1dz\psi^*_+(z)\tilde\psi(z,t)$. Such a Loschmidt echo corresponds actually to the return probability of the evolving $N$-body state $|\tilde{\psi}_+(t)\ra= \sum_n \tilde\psi(z_n,t>t_0)|N-n,n\ra$ to the time-periodic Floquet eigenstate $|\psi_+(t)\ra$,  i.e. ${\cal G}(t>t_0)=\la\psi_+(t)|\tilde{\psi}_+(t)\ra$. The Floquet eigenstate $|\psi_+(t)\ra$ evolves periodically in time with the period $\frac{2\pi}{\omega}$ (modulo time-dependent global phase). The state $|\tilde{\psi}_+(t)\ra$ also reveals nearly periodic behavior on short time intervals around any $t$,  however, it becomes very quickly nearly orthogonal to $|\psi_+(t)\ra$ if $N$ is large. Dynamical quantum phase transitions are associated with non-analytic behavior of the Loschmidt echo $|{\cal G}(t>t_0)|^2$. In order to study the quantum phase transitions in time, it is convenient to define an intensive rate function
\be\label{lambda_plus}
\lambda_+(t)=\lim\limits_{N\rightarrow \infty}\lambda_+^{(N)}(t), \quad \lambda_+^{(N)}= -N^{-1}\ln |{\cal G}(t)|^2 . 
\ee
In Fig.~\ref{Loschmidt}(a) we show the rate function $\lambda_+^{(N)}(t)$ obtained analytically within the continuum approximation, i.e. starting with $\psi(z,t_0) = [\psi_L(z)+\psi_R(z)]/\sqrt{2}$ for $\gamma =-1.15$ and $N=1500$ we obtain subsequent evolution of  $\tilde\psi(z,t>t_0)$ for $\gamma=0$ by harmonic approximation of the potential (\ref{effp}) and by dropping $\sqrt{1-z^2}$ in (\ref{ch}). (See Appendix A for the derivation  of the rate function $\lambda_+(t)$.) The analytical results follow closely the results of the full numerical integration of the $N$-body Schr\"odinger equation (also shown in Fig.~\ref{Loschmidt}) and allow us to obtain $\lambda_+^{(N)}(t)$ in regimes where the Loschmidt echo is so small that the numerical precision breaks down. Cusp-like non-analytic behavior of $\lambda_+(t)$ appear at critical times $t_c=t_0+\frac{\pi}{2J}$ (modulo $\frac{\pi}{J}$). That is, one can show that  $\frac{d}{dt}\lambda_+(t)$ is discontinuous, 
\be 
 \lim\limits_{t\rightarrow t_c^\pm}\frac{d\lambda_+(t)}{dt}=\lim\limits_{t\rightarrow  t_c^\pm}\lim\limits_{N\rightarrow \infty}\frac{d\lambda_+^{(N)}(t)}{dt} =\mp \frac{2q^2\Omega^3}{(1+\Omega^2)^2},
 \ee 
where $\Omega=\sqrt{\gamma(1-\gamma^2)}(1-q^2)^{-1/4}$. It should be stressed that the order of the limits is important. Indeed, $\lim\limits_{t\rightarrow t_c}\lambda_+(t)$ is  well  defined but
 \be
\lim\limits_{N\rightarrow \infty} \lambda_+^{(N)}(t_c) = \frac{q^2\Omega}{1+
 \Omega^2} -\lim_{N\rightarrow \infty}\frac{1}{N} \ln\left[ \cos^2 \left( \frac{ N q^2 \Omega^2}{2(1+ \Omega^2)}\right) \right], 
\label{lambda_tc}
\ee 
is not because whenever the cosine in \eqref{lambda_tc} is close to zero, the logarithm diverges which corresponds to accidental values of $N$ for which the Loschmidt echo vanishes. In Fig.~\ref{Loschmidt}(b) we show $\lambda_+^{(N)}(t)$ around the first critical moment of time for different $N$. With the increasing particle number we get closer to the non-analytical behavior but for some specific values of $N$ the rate diverges.

The Loschmidt echo can be interpreted as information about the evolving system from the point of view of the initial Floquet eigenstate. At the critical time we are not able to extrapolate such information because of the breakdown of a short time expansion. Hilbert space of the system is spanned by the Fock states $|n_1,n_2\ra$ where only two modes $\phi_{1,2}(x,t)$ are occupied by particles.  It turns out that not only the Loschmidt echo but also the von Neumann entropy of the reduced density matrix of the system after the quench, i.e. $S(t>t_0)=-\mbox{tr}\left( \rho_1 \ln \rho_1 \right) $ where $\rho_1=\sum_{n_2}\la n_2|\tilde{\psi}_+(t)\ra\la\tilde{\psi}_+(t)|n_2\ra$, reveals non-analytical behavior at critical moments of times. Indeed, within the continuum approximation $S(t)=-\sum_n |\tilde\psi(z_n,t)|^2\ln  |\tilde\psi(z_n,t)|^2$ and 
\be\label{entanglement_entropy_diff}
\lim\limits_{t\rightarrow t_c^\pm}\left[S(t) -S(t_c) \right] \approx 1.
\ee
Figure~\ref{entropy} illustrates such a sudden jump of the entropy in the system of $N=10^4$ particles, see Appendix B.  

\begin{figure}[bt] 	            
\includegraphics[width=0.65\columnwidth]{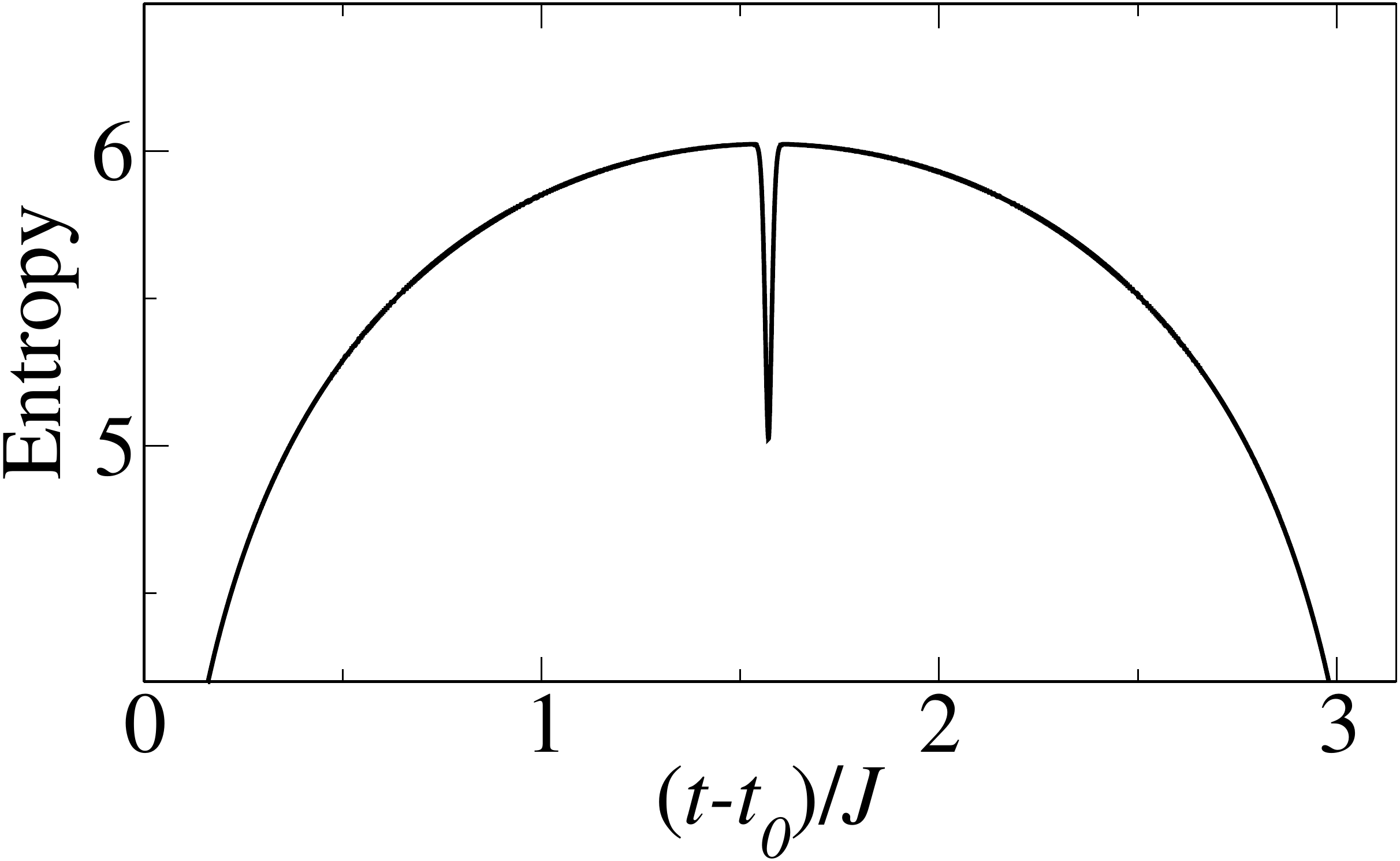}       
\caption{Quench from $\gamma=-50$ to $\gamma=0$ in the system of $N=10^4$ particles prepared initially in the Floquet state $|\psi_+(t)\ra$. Figure shows von Neumann entropy of the reduced density matrix of the system, i.e. $S(t>t_0)=-\mbox{tr}\left( \rho_1 \ln \rho_1 \right)$ where $\rho_1=\sum_{n_2}\la n_2|\tilde{\psi}_+(t)\ra\la\tilde{\psi}_+(t)|n_2\ra$, after the quench. The results are obtained by numerical integration of the $N$-body Schr\"odinger equation. The jump of the entropy visible around the critical moment of time is consistent with Eq.~(\ref{entanglement_entropy_diff}). 
}
\label{entropy}   
\end{figure} 

So far we have considered the system prepared initially in the Floquet state $|\psi_+(t)\ra$ that corresponds to the ground state of the Hamiltonian (\ref{h}) -- within the continuum approximation $\psi_+(z)=[\psi_L(z)+\psi_R(z)]/\sqrt{2}$. The first excited eigenstate of (\ref{h}) can be approximated by $\psi_-(z)=[\psi_L(z)-\psi_R(z)]/\sqrt{2}$ and its eigenenergy becomes degenerate with the ground state energy when $N\rightarrow\infty$. If $\gamma\ll -1$, the Floquet state $|\psi_+(t)\ra$ is actually a Schr\"odinger cat-like state and it could be very difficult to prepare it experimentally because any loss of atoms makes the Schr\"odinger cat collapse to one of the states $|\psi_{L,R}(t)\ra$ \cite{Sacha2015}. 
Assume, that for $\gamma\ll -1$ we have prepared the system in the state $|\psi(t)\ra=|\psi_R(t)\ra=[|\psi_+(t)-|\psi_-(t)\ra]/\sqrt{2}$ which is a symmetry broken state because it evolves with the period twice longer than $\frac{2\pi}{\omega}$ \cite{Sacha2015}. At $t=t_0$ we switch the scattering length $g_0$ to zero. If the ground state level of the initial Hamiltonian is degenerate, the Loschmidt echo is generalized to the return probability of the state after the quench to the ground state manifold \cite{Heyl2013,Zunkovic2016}, i.e. 
\be
{\cal P}(t>t_0)=|{\cal G}_L(t)|^2+|{\cal G}_R(t)|^2
\ee
where ${\cal G}_{L,R}(t)=\la\psi_{L,R}(t)|\tilde{\psi}(t)\ra$. The rate function of the return probability reads
\be\label{rate_sym_broken}
\lambda(t)=-\lim\limits_{N\rightarrow\infty}N^{-1}\ln[{\cal P}(t)]={\rm min}[\lambda_L(t),\lambda_R(t)]
\ee
where 
$
\lambda_{L,R}(t)=\lim\limits_{N\rightarrow\infty}\lambda_{L,R}^{(N)}$,
$\lambda_{L,R}^{(N)}=-N^{-1}\ln|{\cal G}_{L,R}(t)|^2.
$

\begin{figure}[bt] 	            
 \includegraphics[width=0.302\columnwidth]{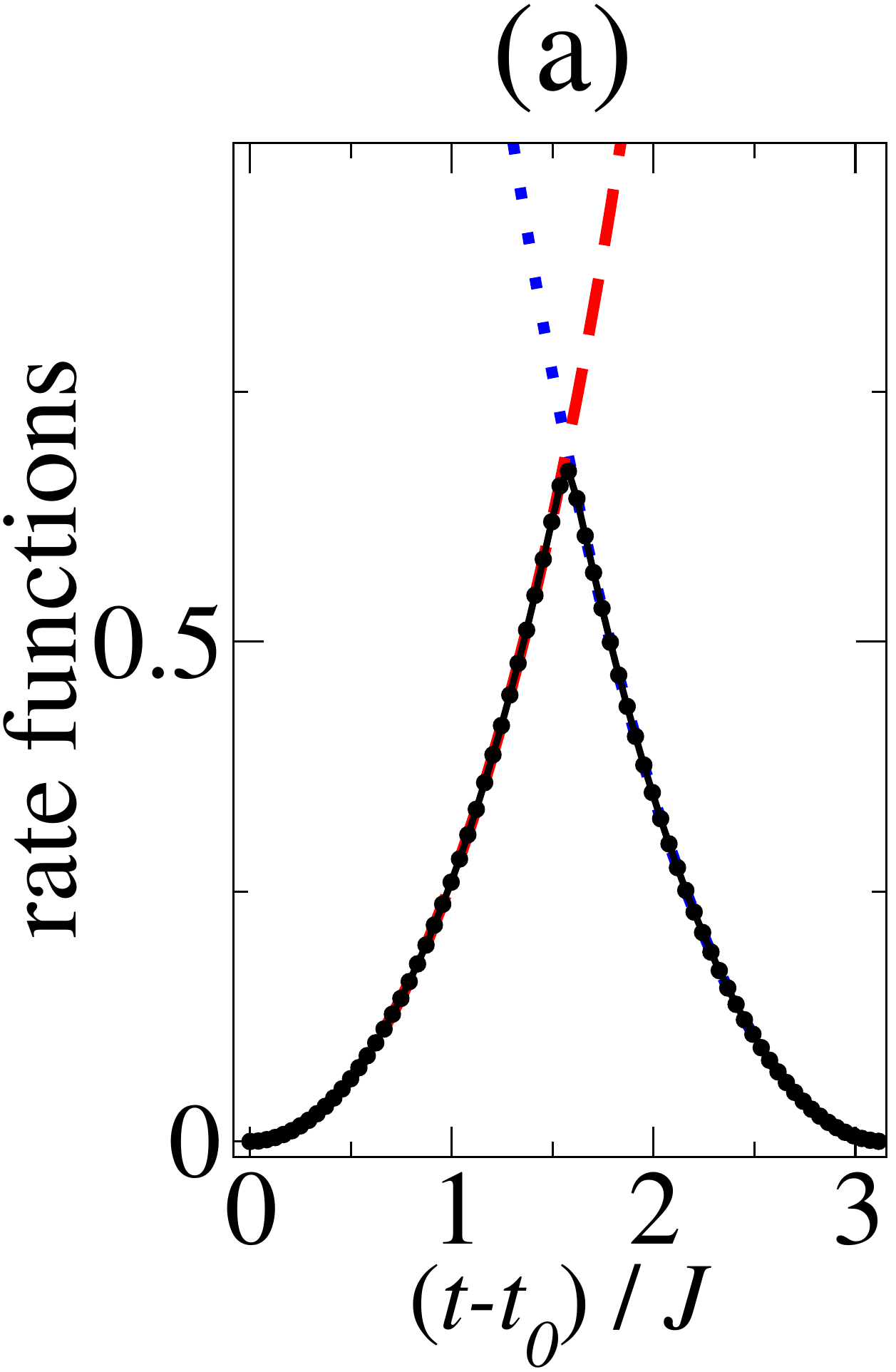}
 \hspace{0.03\columnwidth}
 \includegraphics[width=0.6\columnwidth]{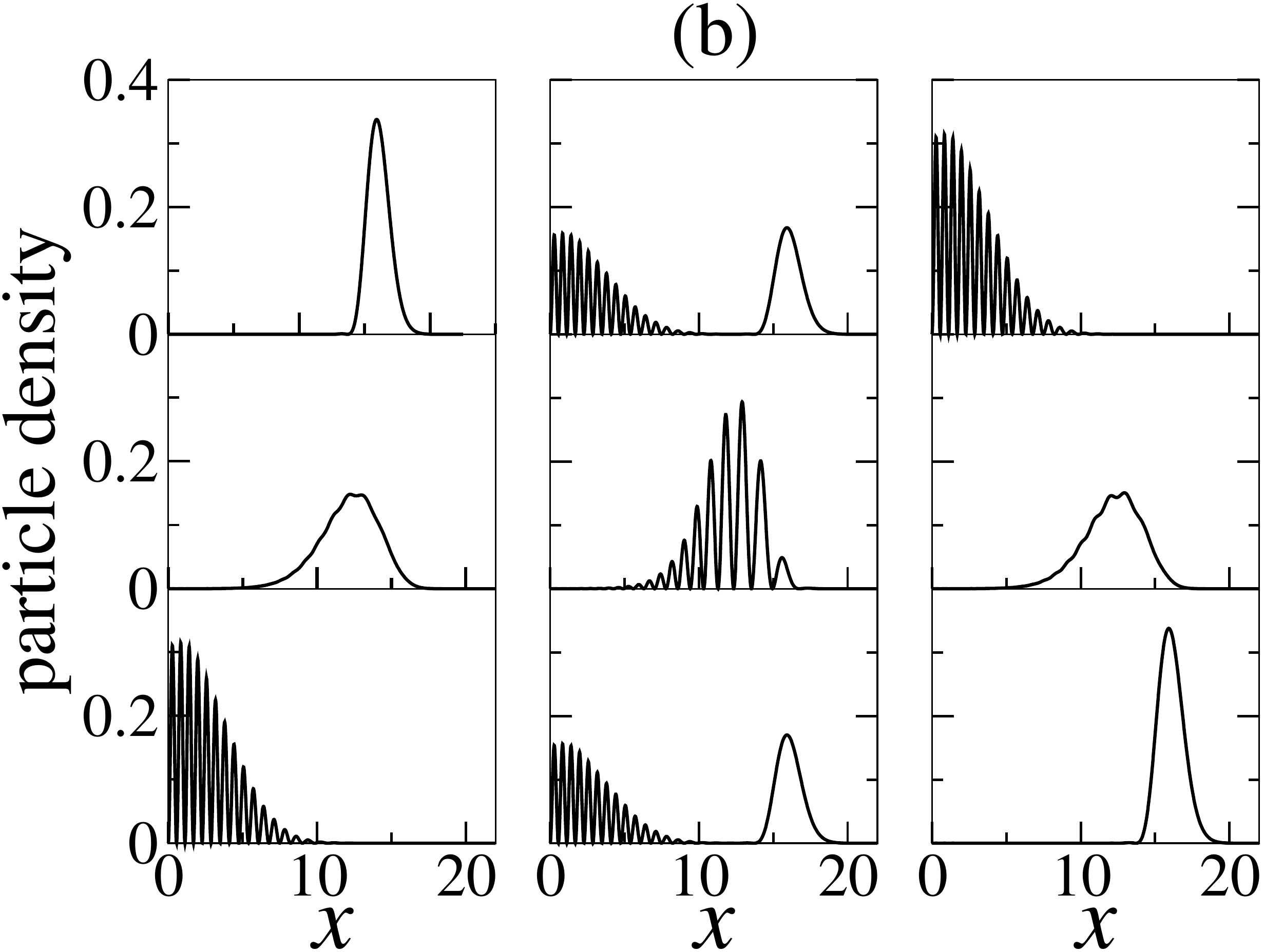}
\caption{
Quench from $\gamma=-50$ to $\gamma=0$ in the system of $N=50$ particles bouncing on a mirror which is located at $x=0$, where $x$ is expressed in the gravitational units. The system is prepared initially in the symmetry broken state $|\psi(t)\ra=|\psi_R(t)\ra$. Panel (a): $\lambda^{(N)}_{R}$ (dashed red line) and  $\lambda^{(N)}_{L}$ (dotted blue line) have been obtained by calculating the projections $|\la\psi_{L,R}(t)|\tilde{\psi}(t)\ra|^2$ while $\lambda$ (black circles) has been determined from $-N^{-1}\ln[{\cal P}(t)]$. The results indicate that even for relatively small particle number $\lambda(t)$ coincides with ${\rm min}[\lambda^{(N)}_L(t),\lambda^{(N)}_R(t)]$. Panel (b): left column illustrates evolution of the density of particles right after the quench, i.e. at $t$ equal $t_0$, $t_0+\frac{\pi}{\omega}$ and $t_0+\frac{2\pi}{\omega}$ from top to bottom, respectively. Middle column presents the particle density at similar time moments but around $t=t_c$ while the right column corresponds to analogues plots around $t=2t_c$. One can see that in the middle column both wavepackets $\phi_{1,2}(x,t)$ are occupied by particles. At moments of time when the wavepackets do not overlap, it is easy to determine their occupations by particles, $|\alpha_{1,2}|^2$, and consequently the rates $\lambda^{(N)}_{L,R}=-\ln|\alpha_{1,2}|^2$.
}
\label{pakiety}   
\end{figure}

Let us stress that the rate \eqref{rate_sym_broken} defined for the initial  symmetry broken state is the same as \eqref{lambda_plus}.  Therefore,  the observed cusps in $\lambda_+(t)$ are a direct consequence of the presence of two symmetry broken states and the fact that the rates $\lambda_{L/R}(t)$ cross at the critical time, see Appendix A.  Let us also stress that although the cusps of Loschmidt echo are difficult to measure, the rates can be easily accessible experimentally. Actually, in an experiment it is not necessary to reach the thermodynamic limit in order to estimate non-analyticities of the rate function $\lambda(t)$. Indeed, if $\lambda_{L,R}^{(N)}(t)$   measured the smallest value of them corresponds to $\lambda(t)$. This procedure has been adopted experimentally \cite{Flaschner2016,Jurcevic2017} and can be also applied in experiments on discrete time crystals. In Fig.~\ref{pakiety}~(a) we show $\lambda_{L,R}^{(N)}(t)$ obtained in the case of relatively small number of particles. The results confirm that $\lambda(t)\approx{\rm min}[\lambda_L^{(N)}(t),\lambda_R^{(N)}(t)]$.
If initially we choose $\gamma\ll -1$, then at $t=t_0$ the system is a Bose-Einstein condensate where all bosons occupy the mode $\phi_2(x,t)$. At  $t=t_c$ we have $\lambda_L(t)=\lambda_R(t)$, and consequently the return probabilities of $|\tilde{\psi}(t)\ra$ to $|\psi_{L,R}(t)\ra$ are equal.  Nevertheless, this does not imply that a state $|\tilde{\psi}(t=t_c)\ra$ is an equal superposition of $|\psi_{L,R}(t=t_c)\ra$. In fact, we find that at all times the state is still a Bose-Einstein condensate with the wavefunction $\alpha_1\phi_1(x,t)+\alpha_2\phi_2(x,t)$. At $t=t_c$ the condensate wavefunction is an equal superposition of $\phi_1(x,t)$ and $\phi_2(x,t)$, and at  $t=t_c+\frac{\pi}{2J}$  the mode $\phi_1(x,t)$ becomes a condensate wavefunction. Fig.~\ref{pakiety}~(b) illustrates evolution of atomic density: right after the quench, around the critical time $t_c=t_0+\frac{\pi}{2J}$ and around $t=t_0+\frac{\pi}{J}$. At moments of time when the wavepackets $\phi_{1,2}(x,t)$ do not overlap the measurement of atomic density allows one to obtain $\alpha_{1,2}$ and consequently the rates $\lambda^{(N)}_{L,R}=-\ln|\alpha_{1,2}|^2$.

\section{Conclusions}

In summary, we have shown that the dynamical quantum phase transitions are not restricted to time independent problems and can also occur in periodically driven systems.  In particular, the dynamical quantum phase transitions can be observed in discrete time crystals.   The Loschmidt echo is related to the return probability of the system after a quench to the initial periodically evolving Floquet eigenstate. We have shown that the first derivative of the rate function of the Loschmidt echo is discontinuous at $t=t_c$.  The cusp in the Lochschmidt echo is a signature of passing the critical point between the discrete time crystal regime and the regime where no spontaneous time translation symmetry breaking can be observed.  Non-analytical behavior of the corresponding rate function has been proven with the help of the so-called continuum approximation where the Hamiltonian of the many-body system is reduced to the one-body Hamiltonian of a fictitious particle.  Dynamical quantum phase transition takes place in the thermodynamic limit which corresponds to the infinite mass of the fictitious particle. The dynamical quantum phase transition described here can be observed in discrete time crystal experiments   if the system is prepared in a state which reveals time translation symmetry breaking.

\section*{APPENDIX A:\\
Time evolution after a quench within the continuum approximation}
\renewcommand{\theequation}{A.\arabic{equation}}
\setcounter{equation}{0}

In the main text we consider a model of a discrete time crystal (DTC), whose description, in the periodically evolving basis, reduces to a $N$-body two-mode Hamiltonian
\bea
\hat H&=&-\frac{J}{2}(\hat a_1^\dagger\hat a_2+\hat a_2^\dagger\hat a_1)+\frac{U}{2}(\hat a_1^\dagger\hat a_1^\dagger\hat a_1\hat a_1+\hat a_2^\dagger\hat a_2^\dagger\hat a_2\hat a_2) \cr
&&+2U_{12}\hat a_1^\dagger \hat a_1\hat a_2^\dagger \hat a_2,
\label{appA:h}
\eea
where  $\hat a_{1,2}$ are standard bosonic  operators which annihilate particles in the periodically evolving modes $\phi_{1,2}(x,t)$, $J$ is hopping amplitude  while  $U$ and $U_{12}$ are interaction strengths between particles in the same and different modes respectively. A relative interaction strength can be quantified by a dimensional parameter $\gamma =N(U-2U_{12})/J$. 

In the thermodynamical limit,  i.e. where $N\rightarrow \infty$ but  $\gamma=$constant, the model has a quantum critial point at $\gamma =-1$ which separates a trivial and a DTC phase. In the  DTC phase, i.e. for $\gamma<-1$,   low-lying eigenstates of (\ref{appA:h}) are vulnerable to  any  perturbation  and spontaneous breaking of the  discrete time translation can be observed \cite{Sacha2015,Sacha2017rev}. 

 Here, we describe the time evolution of the system when one starts with the  DTC regime and  prepares  the system either in the symmetric ground state of \eqref{appA:h} or one of the lowest   symmetry broken states. After the quench through the quantum critical point to the trivial phase (i.e. to $\gamma=0$) we observe singularities of the rate of the Loschmidt echo in time, see the main text.   The after-quench evolution is described analytically   within the so-called continuum approximation \cite{Zin2008}. 

\subsection{Continuous Hamiltonian}

 Let us write the Schr\"odinger equation of the system in the Fock  basis
\be
\la n,N-n | \hat{H} | \psi \ra  = E \la n,N-n  | \psi \ra,
\ee 
where $ | \psi \ra =  \sum_n \psi_n|N-n,n\ra$, or explicitly
\begin{align}
\frac{JN}{2} \left( \psi_{n+1} \sqrt{\frac{1+z_n}{2}\left(\frac{1-z_n}{2}+\frac{1}{N}\right)} \right. + \cr
\left. \psi_{n-1}\sqrt{\frac{1-z_n}{2}\left(\frac{1+z_n}{2}+\frac{1}{N}\right)} -\frac{\gamma}{2} \psi_n z_n^2 \right) =E\psi_n.
\end{align}
For $N\gg 1$ we can
treat the relative population difference $z_n=\frac{(N-n)-n}{N}$ as a continuous variable $z$ with the condition that $|z|\le1$, and $\psi(z_n)=\psi_n$ as continuous wavefunction of a fictitious particle. The continuum approximation reduces the many-body Hamiltonian (\ref{appA:h}) to
\be
\hat H=-\frac{J}{N}\sqrt{1-z^2}\frac{\partial^2}{\partial{z^2}}+V(z), 
\label{appA:ch}
\ee
 where the effective potential reads
\be
V(z)=\frac{JN}{2}\left(-\sqrt{1-z^2}+\frac{\gamma}{2}z^2\right). 
\label{appA:effp}
\ee
 For the latter convenience we set $J=1$.

\subsection{Ground state  manifold }

For $\gamma<-1$ the effective potential $V(z)$ has a double well structure and  the lowest energy eigenstates  of \eqref{appA:ch} can be approximated  by 
\be
\psi_\pm(z) = [\psi_L(z)\pm\psi_R(z)]/\sqrt{2}
\label{appA:sym_gs}
\ee
where $\psi_{L,R}(z)$ are Gaussian states 
 \be
\psi_{L,R}(z)= \frac{1}{(2 \pi \sigma^2)^{1/4}} e^{-\frac{\left(z \pm q \right)^2}{4\sigma^2}},
\ee
 where $q=\sqrt{1-\gamma^{-2}}$, $\sigma=1/\sqrt{N \Omega}$ and $\Omega =\sqrt{\gamma(1-\gamma^2)}/(1-q^2)^{1/4}$. The Gaussian states
are the harmonic oscillator ground states obtained by harmonic expansions of  $V(z)$ around the two local minima $\pm q$
 \be
V(z\approx\pm q) \approx N \tilde{\Omega}^2 (z\mp q)^2/4+ \mbox{constant},
\ee
where $\tilde\Omega = \Omega (1-q^2)^{1/4}$
and by substituting $\sqrt{1-q^2}$ for $\sqrt{1-z^2}$ in (\ref{appA:ch}). For fixed $\gamma$, the larger $N$, the more localized Gaussian states because the total number of particles $N$ is proportional to the mass of the fictitious particle  described by the Hamiltonian (\ref{appA:ch}).

 \subsection{After-quench evolution}

Let us assume that for large but finite $N$ the system is initially prepared in the symmetric ground state  $\psi(z,0)=\psi_+(z)$  \eqref{appA:sym_gs}.
At $t=t_0=0$ we set $\gamma=0$ and describe the time evolution of the system by means of the continuous Hamiltonian \eqref{appA:ch} where the term $\sqrt{1-z^2}$ is neglected. 
\bea
\tilde{\psi}(z,t) &=& \int \mbox{d}z' K(z,t;z') \psi_+(z), \\
K(z,t;z') &=& \sqrt{\frac{a(t)}{\pi i}} e^{i a(t)\left[ \left(z^2+z'^2\right) \cos(t) -2 zz' \right]  }, \cr &&
\eea
where $a(t)=N/\left(4 \sin(t)\right)$. Explicitly  we obtain
\bea
\tilde{\psi}(z,t) &=& [\tilde{\psi}_L(z,t)+\tilde{\psi}_R(z,t)]/\sqrt{2} \, , \label{appA:sym_psi_t}\\
\tilde{\psi}_{L,R}(z,t) &=& A(t) e^{-B(t)z^2 \pm i C(t)z +D(t)} , \label{appA:LR_psi_t}
\eea
where  
\bea
A(t) &=& \sqrt{\frac{-i a(t) }{b(t)\sqrt{2 \pi \sigma^2}}} ,\\ 
B(t) &=& -a(t)\cos(t) +\frac{a(t)^2}{b(t)} ,\\
C(t) &=& \frac{|q|a(t)}{2\sigma^2 b(t) } ,\\
D(t) &=&  \frac{q^2}{4\sigma^2}\left(- 1  + \frac{1}{4\sigma^2 b(t)} \right) ,\\
b(t) &=& \frac{1}{4\sigma^2} - i a(t) \cos(t).
\eea

\subsection{Loschmidt echo and rate function}


In this part we give explicit formulas, within the continuum approximation, for the Loschmidt echo and the rate function \cite{Heyl2017rev} when initially the system is prepared in the symmetric ground state \eqref{appA:sym_gs}.   The initial state is written in the periodically evolving basis and corresponds to the Floquet state that fulfils the discrete time translation symmetry of the original time-dependent Hamiltonian.

The  probability of the return  of the evolving $N$-body state $|\tilde{\psi}_+(t)\ra= \sum_n \tilde\psi(z_n,t>t_0)|N-n,n\ra$ to the time-periodic Floquet symmetric eigenstate $|\psi_+(t)\ra$,  is given by the Loschmidt echo $|{\cal G}(t>t_0)|^2 $, where
\be
{\cal G}(t>t_0)=\la\psi_+(t)|\tilde{\psi}_+(t)\ra \label{appA:l_amplt1}
\ee
is called the Loschmidt amplitude. Within the continuum approximation \eqref{appA:l_amplt1} is equivalent to 
\be
{\cal G}(t)=\int \psi^*_+(z)\tilde\psi(z,t).\label{appA:l_amplt2}
\ee
The Loschmidt amplitude \eqref{appA:l_amplt2} can be written in terms of the symmetry broken states \eqref{appA:sym_gs} and \eqref{appA:sym_psi_t}, namely
\begin{align}
{\cal G}(t)= \int \psi^*_L(z)\tilde\psi_L(z,t)+ \int \psi^*_R(z)\tilde\psi_L(z,t) = 
 \nonumber \\
= \int \psi^*_R(z)\tilde\psi_R(z,t)+ \int \psi^*_L(z)\tilde\psi_R(z,t) =  \nonumber \\
\equiv {\cal G}_R(t)+{\cal G}_L(t),
\end{align}
where, after a conscientious calculation we obtain:
\bea
 {\cal G}_R(t)  &=& \sqrt{\frac{-i a(t) }{2  \sigma^2 b(t)  c(t) }} e^{-\frac{N}{2}\left(\lambda_R(t) + i \mu_R(t)  \right)} , \\
  {\cal G}_L(t)  &=& \sqrt{\frac{-i a(t) }{2  \sigma^2 b(t)  c(t) }} e^{-\frac{N}{2}\left(\lambda_L(t) + i \mu_L(t)  \right)},
\eea
where
\bea
c(t) &=& -i a(t)\cos(t) + \frac{a(t)^2}{b(t)} +\frac{1}{4\sigma^2},
\eea
and 
\bea
\lambda_R(t) &=& \frac{q^2\tilde{\Omega}\tan^2(\frac{t}{2})}{\tilde{\Omega}^2+\tan^2(\frac{t}{2})}, \\
\mu_R(t) &=&  \frac{q^2\tilde{\Omega}^2\tan(\frac{t}{2})}{\tilde{\Omega}^2+\tan^2(\frac{t}{2})}, \\
\lambda_L(t) &=&   \frac{q^2\tilde{\Omega}\cot^2(\frac{t}{2})}{\tilde{\Omega}^2+\cot^2(\frac{t}{2})}, \\
\mu_L(t) &=&  \frac{-\,q^2\tilde{\Omega}^2\cot(\frac{t}{2})}{\tilde{\Omega}^2+\cot^2(\frac{t}{2})}.
\eea

The Loschmidt echo for  non-critical states  scales exponentially with the total number of particles $N$ \cite{Heyl2017b}. Therefore, it is convenient to consider an intensive quantity, i.e. the rate of the Loschmidt echo
\be\label{appA:lambda_plus}
\lambda_+(t)=\lim\limits_{N\rightarrow \infty}\lambda_+^{(N)}(t), \quad \lambda_+^{(N)}= -N^{-1}\ln |{\cal G}(t)|^2 . 
\ee
A straightforward calculations shows that in the thermodynamical limit 
 \be \label{appA:lambda_t}
 \lambda_+(t)=\min \left(\lambda_R(t),\lambda_L(t)\right).
 \ee 
The rate function \eqref{appA:lambda_t} has a non-analytic   cusp for $t=t_c=\pi/2$ when $\lambda_R=\lambda_L$. This   cusp results in discontinuity of the first derivate of the rate function
\be 
 \lim\limits_{t\rightarrow t_c^\pm}\frac{d\lambda_+(t)}{dt}=\lim\limits_{t\rightarrow  t_c^\pm}\lim\limits_{N\rightarrow \infty}\frac{d\lambda_+^{(N)}(t)}{dt} =\mp \frac{2q^2 \Omega^3}{(1+\Omega^2)^2}.
 \ee 
 
 It should be stressed that the order of the limits is important. Indeed, $\lim\limits_{t\rightarrow t_c}\lambda_+(t)$ is defined but
 \be
\lim\limits_{N\rightarrow \infty} \lambda_+^{(N)}(t_c) = \frac{q^2\Omega}{1+
\Omega^2} -\lim_{N\rightarrow \infty}\frac{1}{N} \ln\left[ \cos^2 \left( \frac{ N q^2 \Omega^2}{2(1+\Omega^2)}\right) \right], 
\label{appA:lambda_tc}
\ee 
is not because whenever the cosine in \eqref{appA:lambda_tc} is close to zero, the logarithm diverges which corresponds to accidental values of $N$ for which the Loschmidt echo vanishes.


%
%
%
%

\section*{APPENDIX B:\\
Calculation of entanglement entropy}
\renewcommand{\theequation}{B.\arabic{equation}}
\setcounter{equation}{0}

In this section we present the calculation of the non-analytical jump of the entanglement entropy $S(t)$ at the critical point $t=t_c$ when the system is initially prepared in the symmetric Floquet state $\psi_+(z)$ \eqref{appA:sym_gs}. 

The von Neumann entropy of the reduced density matrix of the system after the quench is defined as
\be
S(t>t_0)=-\mbox{tr}\left( \rho_1 \ln \rho_1 \right),
\ee
where 
\be
\rho_1=\sum_{n_2}\la n_2|\tilde{\psi}_+(t)\ra\la\tilde{\psi}_+(t)|n_2\ra.
\ee
After decomposing the Floquet state in the Fock basis $|\tilde{\psi}_+(t)\ra = \sum_n \tilde{\psi}(z_n,t) |n,N-n\ra$  a straightforward calculation  leads to
\be
S(t)=-\sum_n |\tilde\psi(z_n,t)|^2\ln  |\tilde\psi(z_n,t)|^2 .
\ee

Let us first consider $S(t)$ away from  the critical time. Applying   the continuum approximation we obtain
\begin{align}
S(t\ne t_c) = -\int |\tilde{\psi}|^2 \ln |\tilde{\psi}|^2 \approx \ln 2 \, +  \nonumber \\
-\frac{1}{2}\left( \int |\tilde{\psi}_L|^2 \ln |\tilde{\psi}_L|^2 + 
 \int |\tilde{\psi}_R|^2 \ln |\tilde{\psi}_R|^2 
\right),
\label{appB:entropy1}
\end{align}

where we have  used  the fact  that away from the critical   time $|\tilde{\psi}|^2 \approx [|\tilde{\psi}_L|^2 + |\tilde{\psi}_R|^2]/2$ for $N\gg 1$.
 For  sufficiently large $N$,  \eqref{appB:entropy1} holds for any $t$ arbitrary close to $t_c$. In particular
\begin{align}
\lim_{t\rightarrow t_c^\pm} S(t\ne t_c)\approx \ln 2 
- \int |\tilde{\psi}_0|^2 \ln |\tilde{\psi}_0|^2, 
\end{align}
where , from \eqref{appA:sym_psi_t}-\eqref{appA:LR_psi_t}, we have
\be
|\tilde{\psi}_0|^2 = \lim_{t\rightarrow t_c} |\tilde{\psi}_{L/R}|^2  = \sqrt{\frac{2}{\pi N \Omega}}\; e^{-N z^2/(2\Omega)}.
\ee
 On the other hand, exactly at $t=t_c$ one gets
\begin{align}
S(t_c)= -\int \left|\frac{\tilde{\psi}_0 e^{i\beta z}+\tilde{\psi}_0 e^{-i \beta z}}{\sqrt{2}}\right|^2 \ln \left|\frac{\tilde{\psi}_0 e^{i\beta z}+\tilde{\psi}_0 e^{-i \beta z}}{\sqrt{2}}\right|^2 
 \nonumber \\
 \approx
 -\int |\tilde{\psi}_0|^2 \ln |\tilde{\psi}_0|^2 
 -\int |\tilde{\psi}_0|^2 \cos (2\beta z) \, + \nonumber \\
 -2 \int |\tilde{\psi}_0|^2 \cos^2(\beta z) \ln \left( 1+\cos (2\beta z)\right),
\label{appB:entropy2}
\end{align}
where $\beta=Nq/2$. Note that because of the presence of  $\cos^2(\beta z)$ there are no singular points under the integral in  the second term. Subtracting  \eqref{appB:entropy2} from  \eqref{appB:entropy1} we obtain
\bea
\delta S(t_c) &\equiv& \lim_{t\rightarrow t_c^\pm} S(t\ne t_c)-  S(t_c) \cr
& \approx&  \ln 2 +2\int |\tilde{\psi}_0|^2 \cos (2\beta z) 
\cr && 
+ \sum_{n=2}^{\infty}\int |\tilde{\psi}_0|^2 \frac{(-1)^n}{n(n-1)} 
\cos^n (2\beta z), 
\cr && 
\label{appB:entropy_diff}
\eea
where a series expansion of the logarithm was used. Highly oscillatory terms in \eqref{appB:entropy_diff} can be dropped  if $N\gg 1$.  Note that
\be
\cos^n (2\beta z) = \left(\frac{e^{2i\beta z}+e^{-2i \beta z}}{2}\right)^n =  \sum_{k=0}^n
\left(
\begin{matrix}
 n\\
 k
\end{matrix}
\right)
\frac{e^{2i\beta z (n-2k)}}{2^n} \label{appB:cosine}
\ee
 and that the only  non-oscilatory term in the sum \eqref{appB:cosine} corresponds to $k=n/2$. Using this result in \eqref{appB:entropy_diff} we finally get
\be
\delta S(t_c) \approx \ln 2 + \sum_{m=1}^{\infty} \left(
\begin{matrix}
 2m\\
 m
\end{matrix}
\right)
\frac{1}{2m(2m-1)4^m} = 1 .
\ee

\section*{Acknowledgements}

Support of the National Science Centre, Poland via Projects No. 2016/21/B/ST2/01086 (A.K.) and No. 2016/21/B/ST2/01095 (K.S.) is acknowledged.


\end{document}